\documentclass[10pt,journal,letterpaper,compsoc,twoside]{IEEEtran}
\ifCLASSOPTIONcompsoc
  \usepackage[nocompress]{cite}
\else
\fi
\usepackage[dvips]{graphicx}
\usepackage{balance}

\usepackage[cmex10]{amsmath}
\usepackage{amssymb}
\usepackage{array}

\ifCLASSOPTIONcompsoc
\usepackage[tight,normalsize,sf,SF]{subfigure}
\else
\usepackage[tight,footnotesize]{subfigure}
\fi

\ifCLASSOPTIONcompsoc
  \usepackage[font=normalsize,labelfont=sf,textfont=sf]{subfig}
\else
  \usepackage[font=footnotesize]{subfig}
\fi
\usepackage{url}

\usepackage{multirow}
\begin{document}
%
\title{A High-Performance Solid-State Disk with Double-Data-Rate NAND Flash Memory}
%
%
%
%

\author{Eui-Young~Chung,~\IEEEmembership{Member,~IEEE}, Chang-Il~Son, Kwanhu~Bang,~\IEEEmembership{Student~Member,~IEEE,}\\Dong~Kim,  Soong-Mann Shin,
        and~Sungroh~Yoon,~\IEEEmembership{Senior Member,~IEEE}
\IEEEcompsocitemizethanks{
\IEEEcompsocthanksitem  Eui-Young Chung and Kwanhu Bang are with School of Electrical and Electronic Engineering, Yonsei University, Seoul, Korea. E-mail: \{eychung, lamar49\}@yonsei.ac.kr
\IEEEcompsocthanksitem  Chang-Il~Son, Dong~Kim and Soong-Mann Shin are with Flash Solution R\&D Center, Samsung Electronics, Hwasung, Kyungki, Korea.
E-mail: \{cison, dong.kim, sm1978.shin\}@samsung.com
\IEEEcompsocthanksitem Sungroh Yoon (corresponding author) is with Department of Electrical and Computer Engineering, Seoul National University, Seoul, Korea. E-mail: sryoon@snu.ac.kr}
\thanks{Manuscript drated June 4, 2009.}
}


%
%

\markboth{DRAFT}%
{Chung \MakeLowercase{\textit{et al.}}: A High-Performance Solid-State Disk with Double-Data-Rate NAND Flash Memory}
%



\IEEEcompsoctitleabstractindextext{%
\begin{abstract}
We propose a novel solid-state disk (SSD) architecture that utilizes a double-data-rate synchronous NAND flash interface for improving read and write performance. Unlike the conventional design, the data transfer rate in the proposed design is doubled in harmony with synchronous signaling. The new architecture does not require any extra pins with respect to the conventional architecture, thereby guaranteeing backward compatibility. For performance evaluation, we simulated various SSD designs that adopt the proposed architecture and measured their performance in terms of read/write bandwidths and energy consumption. Both NAND flash cell types, namely single-level cells (SLCs) and multi-level cells (MLCs), were considered. In the experiments using SLC-type NAND flash chips, the read and write speeds of the proposed architecture were 1.65--2.76 times and 1.09--2.45 times faster than those of the conventional architecture, respectively. Similar improvements were observed for the MLC-based architectures tested. It was particularly effective to combine the proposed architecture with the way-interleaving technique that multiplexes the data channel between the controller and each flash chip. For a reasonably high degree of way interleaving, the read/write performance and the energy consumption of our approach were notably better than those of the conventional design.
\end{abstract}

}

\maketitle

\IEEEdisplaynotcompsoctitleabstractindextext

%
\IEEEpeerreviewmaketitle

\section{Introduction} \label{s-intro}
%
%
%
%

\IEEEPARstart{N}{AND}-flash-based solid-state disks (SSDs) are replacing hard disk drives (HDDs), the mass storage device of choice for many decades, not only in high-end servers but also in mainstream PCs and in low-end mobile internet devices (MIDs). The compelling reason for this change can be attributed to the absence of mechanical moving parts in SSDs; this fact can substantially enhance key characteristics of mass storage devices such as read/write performance, power consumption, weights, form factors, reliability, shock resistance, and many others.
In particular, the improved read/write performance of SSDs is expected to narrow the so-called \emph{CPU-IO performance gap}~\cite{katz1989dsa}, which has been a long-standing problem for accelerating computer systems. Due to the recent advent of multi-core CPUs, the CPU-IO performance gap would become even wider without a breakthrough in IO systems. Thus, the read/write performance has become one of the most important metrics to determine the overall merit of a storage device.


The two major components of a typical NAND-based SSD are the following: i) a number of NAND flash memory chips and ii) a control circuitry called the \emph{SSD controller}, which manages the data transfer between the NAND flash chips and the host machine the SSD is attached to.
%
%
%
The system-level read/write speed of an SSD is often orders of magnitude faster than that of HDDs, but this is not because the individual NAND flash chips inside the SSD are that fast. In fact, a major performance bottleneck in an SSD may occur due to the latency of accessing NAND flash memory.
For instance, the time to program a flash cell
is normally in the range of hundreds of microseconds, which is several orders of magnitude greater than the typical clock-cycle time of the SSD controller.
Thus, the SSD controller should frequently slow down or be idle in order
to keep pace with NAND flash memory, thereby incurring a performance loss.
%
SSDs can be faster than HDDs because of the various techniques employed to hide and/or reduce the latency of sluggish NAND flash memory, as will be surveyed shortly.

The NAND flash access time issue has become more critical due to the advent of multi-level-cell (MLC) flash memory. A traditional NAND flash chip can store only one bit per cell and is called single-level-cell (SLC) flash memory. In contrast, MLC flash memory can store multiple bits per cell. Thus, MLC flash memory is more cost-effective, since it demands much less die space than SLC flash memory, in order to integrate the same capacity using the same process technology.  Unfortunately, the MLC implementation inevitably increases the access time. For instance, it is known that the cell program time of
MLC flash memory is approximately three times larger than that of SLC flash memory.
Nevertheless, the adoption of MLC flash memory will rapidly grow, since the MLC implementation can significantly lower the per-bit cost, which is still much higher than that of HDDs.

%

The core competency of SSDs over the HDDs can thus be obtained by trading off the access time and the cost of NAND flash memory in an effective manner. This point was recognized early, and many techniques have been proposed to alleviate the access time issue. As detailed in Section~\ref{s-related}, examples include way interleaving, channel striping, and caching. Strictly speaking, these techniques are more for hiding the NAND flash access latency, rather than reducing it. There exist other approaches targeting on actual reduction of the latency. A key idea of these techniques is to replace the conventional asynchronous NAND flash interface scheme by a synchronous one, an idea that stems from the history of DRAM: the initial asynchronous DRAM interface was later replaced by faster synchronous interfaces. However, the limitation of these approaches is that they require additional pins, thereby causing area overhead and incompatibility with the traditional components.



Our approach proposed in this paper belongs to the category of techniques to reduce the latency itself. More precisely, the contributions of our work are two-fold. First, we propose a novel SSD architecture that utilizes a double-data-rate (DDR) synchronous NAND flash interface for improving read and write performance. Unlike the conventional design, the data transfer rate in the proposed design is doubled in harmony with synchronous signaling. Furthermore, the new architecture does not require any extra pins with respect to the conventional architecture, thereby guaranteeing backward compatibility. Second, we thoroughly validate the performance of our approach by simulating various SSD designs that adopt the proposed architecture and by measuring their read and write bandwidths as well as energy consumption. Moreover, we show how the proposed architecture is combined with the two most popular latency-hiding techniques, namely way interleaving and channel striping, for their synergistic effects on overall performance at SSD-level. For realistic results, we consider both SLC and MLC NAND flash memory.

%
%


The rest of this paper is organized as follows. Section~\ref{s-preliminary} introduces the basics of SSD architectures and discusses possible options for enhancing SSD performance. This section also provides a brief review on previous approaches for resolving the latency issue in NAND flash memory. In Section~\ref{s-conventional}, we describe the conventional SSD architecture that uses the single-date-rate asynchronous NAND flash interface. The proposed SSD architecture that utilizes the new DDR synchronous NAND flash interface is detailed in Section~\ref{s-ddr}. Finally, we provide our experimental results in Section~\ref{s-experiment} followed by a conclusion in Section~\ref{s-conclusion}.


\section{Preliminaries and Related Work} \label{s-preliminary}
\subsection{Typical SSD Architecture}
\begin{figure}[!t]
\centering
\includegraphics[width=\linewidth]{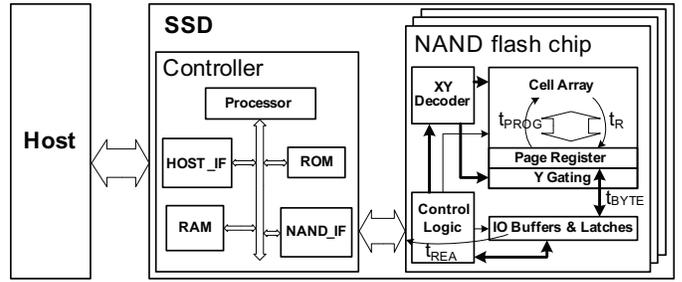}
\caption{Block diagram of a typical SSD.} \label{fig1}
\end{figure}

Fig.~\ref{fig1} shows the architecture of a typical SSD, which is composed of multiple NAND flash memory chips and a controller to manage the data transfer between the host machine and the NAND flash chips.
The controller contains various components such as a processor, random access memory (RAM), read only memory (ROM), a host interface, and a NAND interface. The processor governs the controller by executing the firmware residing in the ROM chip. Some notable tasks of the processor include wear leveling and address translation, as will be explained in Section~\ref{ss-prelim-control}. The NAND interface labeled NAND\_IF in Fig.~\ref{fig1} is to communicate with the NAND flash chips.

Each NAND flash memory chip in the SSD architecture is composed of a cell array, a page register, an XY decoder, a control logic, IO buffers, and latches. The cell array stores the entire set of data, while the page register temporarily stores one page of the data being requested for read or write. The XY decoder decodes the address issued by the controller, and the
control logic manages the interface with the controller. The data
transfer time from the cell array to the page register is defined
as $t_R$, and the time for the reverse action (i.e. the time to transfer data from the page register to the cell array) is called the page program time or $t_{PROG}$. Typically, $t_{PROG}$ is much larger than $t_R$. The data transfer time between the page register and IO buffer is referred to as $t_{BYTE}$. Finally, $t_{REA}$ is the data transfer time between
the IO buffer and IO pads. More details on these timing parameters will be presented in Table~\ref{t-timing-proposed} in Section~\ref{s-conventional}.

\subsection{Options for Improving SSD Performance}

For the SSD architecture shown in Fig.~\ref{fig1}, the opportunities for performance improvements can be summarized as follows:
i) to enhance the performance of NAND flash cells, ii) to optimize the performance of the SSD controller, iii) to use a faster interface between the SSD and its host, iv) to accelerate the interface between the SSD controller and the NAND flash chips, and v) the mixture of these options.
We survey the techniques based on options ii)--iv). Option i) is beyond the scope of this paper and will not be discussed further; interested readers are directed to \cite{r1,r2,r3,r4,r5,r6,r7}. 

\subsubsection{Optimizing SSD Controller}\label{ss-prelim-control}
This may be the option that has been most actively studied. From the hardware perspective, one of the most frequently used techniques is to increase data throughput by parallelizing the data paths between the controller and NAND flash chips. Such paths are called \emph{channels}, and there are largely two methods for the parallelization. One is called \emph{channel striping}, which means using multiple channels in the NAND flash interface. The other is called \emph{way interleaving}, and this is to multiplex each channel to send data in a round-robin fashion. By exploiting these techniques, it is possible to hide much of the latency of NAND flash memory.

\begin{figure}
\centering
\includegraphics[width=0.9\linewidth]{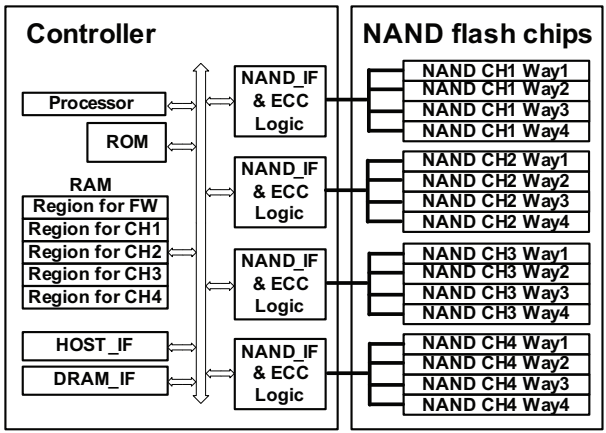}
\caption{An SSD architecture with 4 channels and 4
ways per channel.} \label{fig6}
\end{figure}

Fig.~\ref{fig6} is an example of the SSD architecture adopting the techniques of channel striping and way interleaving simultaneously. The number of channels and ways in this example are both four. Of note is that channel striping is often more costly than way interleaving, since each channel requires a NAND interface block and an error correction code (ECC) block. The ECC block is essential for data reliability, especially when the MLC flash is used. Another area penalty of multi-channel design comes from increased pin counts. Each channel requires dedicated pins to communicate with the dedicated NAND flash memory chips. For this reason, the number of channels should carefully be selected in order to achieve the required system performance within the area budget.




Another performance improvement technique from the controller perspective is to optimize the software called \emph{flash translation layer} (FTL)~\cite{r11,r12,r13}. FTL runs on the processor of an SSD controller and performs mapping between logical and physical addresses and also handles important housekeeping tasks such as wear leveling~\cite{bragdon1995fmm} and garbage collection. Wear leveling is to use all the flash cells in a chip as uniformly as possible and plays a critical role to maintain the initial performance and capacity of an SSD over time, since the lifetime of a flash cell is directly limited by its write frequencies.



Besides, in most commercially available SSDs, DRAM is used as a cache buffer to hide the long access latency of NAND flash memory. If the data requested by the host machine happens to be found in the cache buffer, we can completely eliminate the data access time to NAND flash memory.

Refer to Sections~\ref{ss-related-hiding} and~\ref{ss-related-controller} for a brief survey of the existing approaches that belong to this category.

\subsubsection{Improving Host Interface}
This option is to increase the bandwidth between the SSD and its host machine. Currently, SSDs are attached to the host machine via legacy interfaces inherited from HDDs such as parallel advanced technology attachment (PATA) and serial-ATA  (SATA)~\cite{sata}. To achieve higher performance with less pin counts, SATA is rapidly replacing PATA these days both for HDDs and SSDs. In addition, to handle properly the increased bandwidth of SSDs, alternative high-speed interfaces such as peripheral component interconnect express (PCIe) have been tried for interfacing SSDs. Recently, it was proposed in ~\cite{kim2001ssd} to attach SSDs to the North Bridge chipset using the DRAM interface, instead of using the South Bridge chipset in which the SATA and PATA controllers reside.

\subsubsection{Accelerating NAND Flash Interface}

This is to increase the bandwidth between the controller and each NAND flash memory chip. Even though the objective of this option is similar to that of channel striping or way interleaving, this option is more aggressive in the sense that the read and write bandwidths can be improved by reducing the latency directly, rather than hiding it. A key technique in this category is to improve the NAND flash interface scheme in a synchronous fashion. Section~\ref{ss-related-interface} presents more details of existing techniques for accelerating NAND flash interfaces.

%



\subsection{Related Work} \label{s-related}

\subsubsection{Hiding the Latency of NAND Flash Memory}\label{ss-related-hiding}

The effect of channel striping and way interleaving was extensively
studied in~\cite{r16}, which used a 2-channel, 4-way-interleaving
interface scheme with a software architecture adopting a hybrid-mapping
algorithm. The proposed system outperformed the compared HDD by 77\%.
The improvement was mainly due to the increased parallelism and the
interleaved accesses when programming NAND flash memory. However, the limitations of this approach include area overhead and complicated controller design due to the increased number of channels.
Other approaches to latency-hiding include the techniques proposed in~\cite{cache1, cache2}, where DRAM was used as the cache buffer for NAND flash memory. When a cache hit occurs, the data access time is solely determined by the DRAM access time, which is much smaller than the flash access time.

\subsubsection{Optimizing the Firmware of SSD Controller}\label{ss-related-controller}
The techniques in this category aim at enhancing the
SSD performance by reducing the data transfer
size, operating time, and the number of extra operations
required for wear leveling. The technique presented in ~\cite{r8,r9,r10}
compresses the data from the host unit to save the storage space in NAND flash memory and to reduce the data transfer time from the controller to flash chips. However, this method may incur extra time and area overheads for data (de)compression. The hybrid-mapping technique proposed in~\cite{r11} aimed at improving the write speed by introducing two types of logical blocks called \emph{data blocks} and \emph{log blocks}. The number of log blocks is much smaller than that of data blocks, and data is always written to log blocks first. When all log blocks are used up, the FTL moves the data from log blocks to data blocks. This technique may incur extra computation overhead but can be beneficial for quick search owing to the small number of log blocks. The techniques introduced in~\cite{r12,r13,r14,r15} can reduce the
number of erase operations by using a page-map cache and smart
mapping strategies; it was shown that the system performance can be enhanced by reducing the number of erase and garbage collection operations.

\subsubsection{Improving Controller-Flash Interface}\label{ss-related-interface}
In~\cite{r17}, the authors introduced a synchronous NAND flash interface
using a signal called \emph{data valid strobe} (DVS). This interface
improved the sensitivity to the process, voltage, and
temperature (PVT) as well as the read performance by isolating the
timing of the controller from that of the NAND flash memory. However, this approach exploited only one edge of each clock signal, producing limited performance improvements. The focus of this work was more on desensitizing PVT variations rather than on boosting read and write performance.

Recently, some leading companies in the SSD business organized an initiative called \emph{open NAND flash interface} (ONFI) and proposed a DDR flash interface scheme, whose specification is available at~\cite{onfi}. Additionally, the authors in~\cite{hland} proposed a similar concept along with a new SSD architecture. However, these approaches require additional pins, thus causing compatibility issues and area overhead. Furthermore, no quantitative analysis was performed to prove the effectiveness of these approaches and to show the impact of DDR interface schemes on the SSD performance.

Our work presented in this paper belongs to the category of techniques to accelerate the interface between SSD controller and NAND flash chips. Unlike the aforementioned approaches, our DDR synchronous interface scheme provides pin-level compatibility with the traditional NAND flash memory interface. Moreover, we evaluate the effect of the proposed technique quantitatively with respect to various architectural choices (e.g. the number of channels and ways) from the SSD perspective.

\begin{figure*}
\centering
\includegraphics[width=0.7\linewidth]{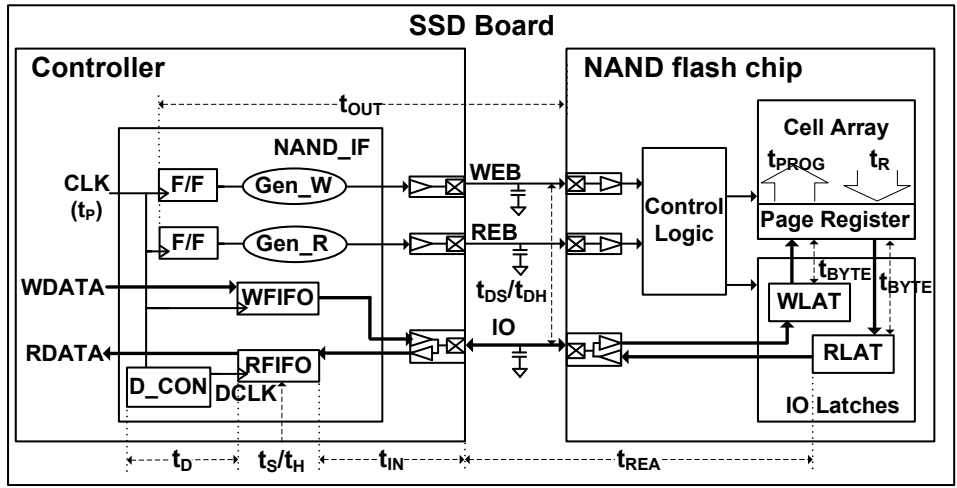}
\caption{Block diagram of the NAND flash memory interface in the conventional SSD architecture.}
\label{fig2}
\end{figure*}

\section{Conventional Asynchronous NAND Flash Interface for Solid-State Disks} \label{s-conventional}
The overall structure of a typical SSD was explained in Section~\ref{s-preliminary}. In this section, we present additional details on the conventional method for interfacing the controller and the NAND flash chips in SSDs. The material in this section is crucial for understanding the new interface architecture proposed in Section~\ref{s-ddr}. The major difference between the two architectures lies in the controller-flash interface; the conventional interface uses a asynchronous single-data-rate scheme, whereas the proposed design utilizes a synchronous double-data-rate scheme.


\subsection{Block Diagram and Key Components}

Fig.~\ref{fig2} shows the conventional asynchronous interface architecture. Note that only the NAND\_IF block is drawn inside the controller block for clarity, although there exist additional blocks, as shown in Fig.~\ref{fig1}.
The NAND\_IF block and the NAND flash chip communicate over three types of ports. The upper two ports are for transferring data strobe signals, and the lower one is for exchanging all the other control signals as well as data.

Inside the NAND\_IF block, there are two blocks called \emph{generate write} (Gen\_W) and \emph{generate read} (Gen\_R). The signal to control writes is called \emph{write enable bar} (WEB) and is generated by the Gen\_W block. The read control signal is named \emph{read enable bar} (REB) and is produced by the Gen\_W block. WEB and REB are sent to the NAND flash chip via the upper two ports of the interface. The D\_CON block is to delay the clock (CLK) so that data transfers at the interface can fulfill any given timing specifications. The blocks called \emph{WFIFO} and \emph{RFIFO} are for buffering data from and to the host, respectively.

The IO latches inside the flash chip include timing-critical parts called \emph{write latch} (WLAT) and \emph{read latch} (RLAT). WLAT temporarily stores the data from the controller to the page register, whereas RLAT temporarily stores the data from the page register to the controller.


%



\subsection{Timing Parameters}

\begin{table*}
\renewcommand{\arraystretch}{1.3}
\centering
\caption{Timing parameters for the conventional and proposed interface architectures.}
\label{t-timing-proposed}
\begin{tabular}{|c||c|c|}
\hline
Parameter & Conventional (Fig.~\ref{fig2})& Proposed (Fig.~\ref{fig4})\\\hline\hline
$t_P$       &\multicolumn{2}{c|}{Clock (CLK) period}\\\hline
$t_{D}$     &\multicolumn{2}{c|}{Delay amount of CLK by D\_CON (i.e. difference between CLK and DCLK); $t_D = \alpha \cdot t_P$, where $0 \le \alpha \le 1/2$}\\\hline
$t_{S}$/$t_{H}$ &\multicolumn{2}{c|}{Setup/hold time of WFIFO and RFIFO}\\\hline
$t_{R}$         &\multicolumn{2}{c|}{Data fetch time (from Cell Array to Page Register)}\\\hline
$t_{PROG}$      &\multicolumn{2}{c|}{Program time (from Page Register to Cell Array)}\\\hline
$t_{BYTE}$ &\multicolumn{2}{c|}{Data transfer time between Page Register and WLAT/RLAT}\\\hline
$t_{WC}$ &\multicolumn{2}{c|}{Write cycle time (i.e. one cycle of WEB)}\\\hline
$t_{RC}$ &\multicolumn{2}{c|}{Read cycle time (i.e. one cycle of REB)}\\\hline
$t_{IN}$   &Data propagation time between the IO pad&\\
            &of the controller and WFIFO/RFIFO&\\\cline{1-2}
$t_{OUT}$   &Signal propagation time from FFs of the controller&\\
            & to the strobe pads of NAND flash memory&N/A\\\cline{1-2}
$t_{DS}$/$t_{DH}$   &Setup/hold time of IO signals with respect to WEB&\\\cline{1-2}
$t_{REA}$           &Data transfer time from RLAT to&\\
                    &the IO pad of the controller&\\\hline
$t_{DIFF}$  &&Difference between the arrival time of DVS at RFIFO\\
            &&and the arrival time of IO in the NAND flash at RFIFO\\\cline{1-1}\cline{3-3}
$t_{DLL}$   &&Time delay by DLL as defined in Eq.~(\ref{eq1})\\\cline{1-1}\cline{3-3}
$t_{RWEBD}$&&Propagation delay of RWEB  from\\
                        &N/A&the strobe port of NAND flash memory to DLL\\\cline{1-1}\cline{3-3}
$t_{IOS}$/$t_{IOH}$   &&Setup/hold time of IO signals with respect to DVS\\\cline{1-1}\cline{3-3}
$t_{IOD}$   &&Data propagation delay from RLAT to the IO pad\\
            &&of NAND flash memory\\\cline{1-1}\cline{3-3}
$t_{RWC}$ &&One cycle of RWEB; replaces $t_{RC}$ and $t_{WC}$\\\hline
\end{tabular}
\end{table*}

To explain the write and read operations of the SSD interface architecture in Sections~\ref{ss-conv-writing} and~\ref{ss-conv-reading}, we first show in Table~\ref{t-timing-proposed} a number of important timing parameters for the interface building blocks. In the table, note that the first eight parameters are common for the conventional and the proposed interfaces. The next four are only for the conventional architecture; the rest are only for the proposed architecture detailed in Section~\ref{s-ddr}. Additional timing parameters of NAND flash chips themselves are available in~\cite{r18,r19,r20}.



\subsection{Write Operation and Timing}\label{ss-conv-writing}

\begin{figure*}
\centering
\subfigure[Write mode] {
    \includegraphics[width=0.6\linewidth]{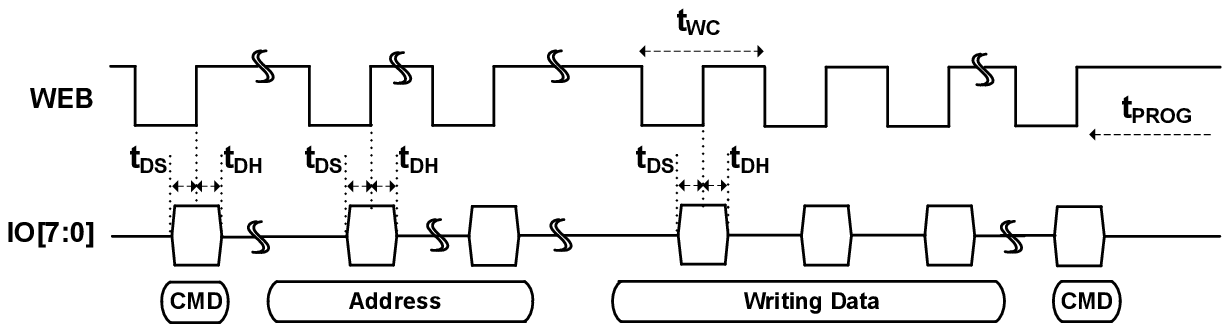}
}
\subfigure[Read mode] {
    \includegraphics[width=0.6\linewidth]{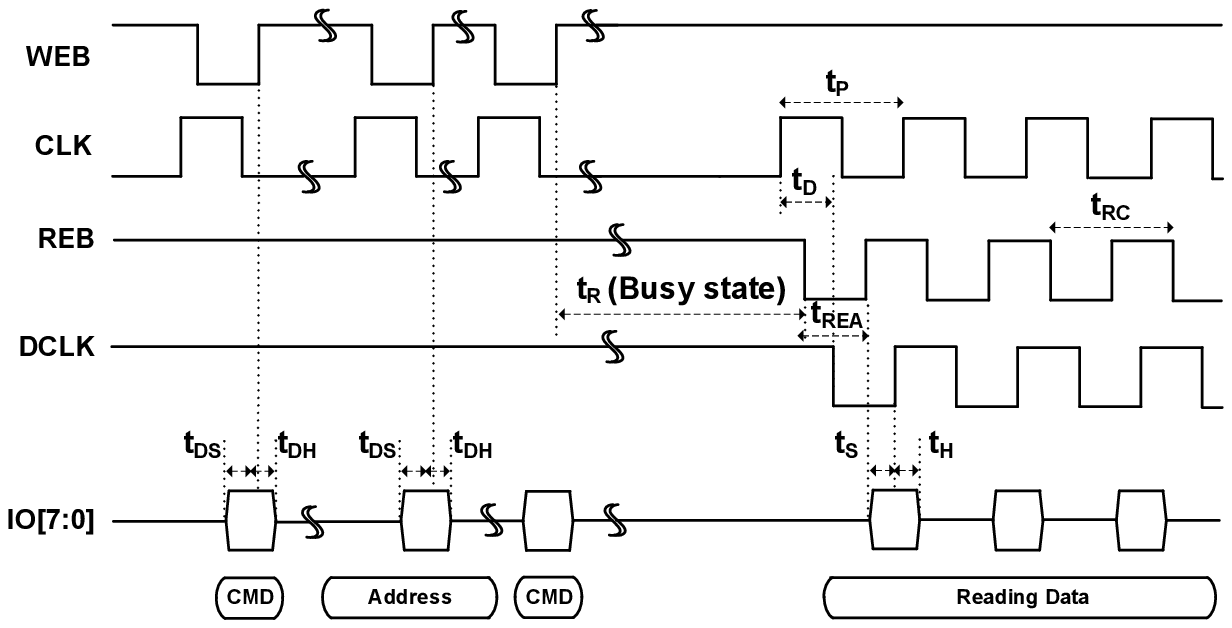}
}
\caption{The timing diagrams of the conventional asynchronous NAND flash memory interface.} \label{fig3}
\end{figure*}

Fig.~\ref{fig3}(a) shows the write timing diagrams of the conventional NAND flash memory interface. The controller asserts WEB and issues the first write command (CMD) to the flash chip in order to initiate a write operation. The destination addresses are then sent to the flash chip followed by a series of data to be written to the page register through WLAT at every $t_{WC}$, the period of WEB. Finally, the controller issues a program CMD to transfer the data in the page register to the cell arrays of the flash chip. During the program phase, the flash memory chip enters the busy state and cannot be interrupted until the end of the program phase. This time duration is defined as $t_{PROG}$ and is normally very long. 

Note that, in the write mode, both control (i.e. WEB) and data are concurrently transferred from the controller to the flash chip; the delays of the control and data paths are almost identical. The conventional interface operates synchronously in the write mode in the sense that transfers are synchronized to the periodic WEB signal under the timing constraints set by $t_{DS}$ and $t_{DH}$. The data transfer rate in the write mode can therefore be improved by increasing the frequency of WEB. However, the conventional interface is not considered synchronous due to the asynchronous read mode, as will be explained next.

\subsection{Read Operation and Timing}\label{ss-conv-reading}


The timing diagrams for the read operation are shown in Fig.~\ref{fig3} (b). After issuing the first read CMD followed by the destination address, the second read CMD is issued to the flash chip. It then enters the busy state for fetching data from the cell arrays to the page register. This data fetching time is defined as $t_R$, which is much shorter than $t_{PROG}$. Thus, the data transfer time between the cell arrays and the page register is not as critical in the read mode as it was in the write mode. At the completion of the fetch, the flash chip enters the ready state, and the controller periodically asserts REB to the flash chip with the period of $t_{RC}$. For each REB cycle, the control logic inside the flash chip instructs a single data transfer from the page register to RLAT within $t_{BYTE}$, and the data reach the IO ports of the controller within $t_{REA}$. The controller then fetches the data into RFIFO at the positive edge of DCLK, a delayed version of CLK by $t_D$. More precisely, $t_D$ is defined as
\begin{equation}\label{e-td}
t_D=\alpha \cdot t_P,
\end{equation}
where $0 \le \alpha \le \frac{1}{2}$. Note that DCLK is used to satisfy the setup time constraint imposed on RFIFO. Without DCLK, the system may easily violate the timing constraint due to the variations of t$_{IN}$, t$_{OUT}$, and t$_{REA}$. Thus, each operation of propagating REB and fetching data is allowed to take at most $t_{RC} + t_D$, instead of $t_{RC}$.

It is critical to notice the following: In the read mode of the conventional interface, the control (i.e. REB) and data cannot be propagated concurrently, unlike the write mode. That is, REB is first propagated from the controller to the flash chip, and then the data transfer occurs in the opposite direction. Consequently, a single read cycle should be determined by the sum of the propagation delays of REB and data, unlike the write mode in which a write cycle can be set by the maximum of the two delays. For this reason, $t_{RC}$ is normally longer than $t_{WC}$, although the specification of commercial NAND flash memory usually lists identical timing parameters for convenience.
The new interface architecture proposed in the next section focuses on reducing the read cycle time in order to enhance read performance.





\section{Proposed DDR Synchronous NAND Flash Interface for Solid-State Disks} \label{s-ddr}
\begin{figure*}
\centering
\includegraphics[width=0.7\linewidth]{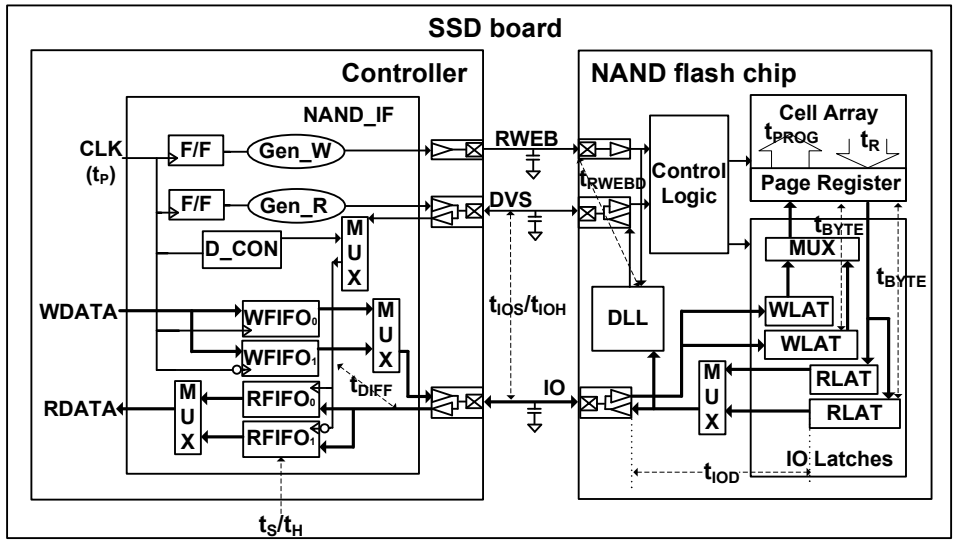}
\caption{Block diagram of the proposed double-data-rate synchronous NAND flash memory interface.}
\label{fig4}
\end{figure*}

In this section, we provide the details of the proposed NAND flash interface for improving SSD performance. This new architecture focuses on enhancing the data throughput between NAND flash memory chips and the SSD controller. To this end, the proposed scheme operates in a synchronous manner for both read and write modes and supports double-data-rate transfers.

As highlighted in Section~\ref{s-conventional}, a major performance bottleneck in the conventional NAND flash memory is the serialized, opposite-directional propagation of control and data in the read mode. The propose interface breaks this serialized propagation paths into two smaller ones --- one for control and the other for data --- and isolate them from the perspective of timing. More precisely, the REB control is generated by CLK and is propagated just as in the conventional architecture. On the other hand, the data is fetched from the flash chip to the controller in synchronization with a new control signal named \emph{data valid strobe} (DVS), as depicted in Fig.~\ref{fig4}. DVS is a data strobe asserted by the flash chip and can be considered as a data clock whose edges indicate stable points for data fetching.

Introducing DVS is for the synchronous read operation. To support DDR operation, we duplicate the RFIFO and WFIFO buffers inside the controller and the RLAT and WLAT latches inside the flash chip. In the controller, one pair of RFIFO and WFIFO is dedicated to the rising edge of CLK, and the other pair to the falling edge of CLK; in the flash chip, one pair of RLAT and WLAT is for the rising edge of DVS, and the other pair for the falling edge for DVS.



The notion of DVS was first introduced in~\cite{r17}, but the purpose of that work was not to increase the data bandwidth but to desensitize the PVT variations as discussed in Section~\ref{s-related}. In contrast to~\cite{r17}, the proposed design can enhance the overall read/write performance of an SSD by allowing double-data-rate data transfers between the controller and flash memory. We compare the performance of the interface introduced in~\cite{r17} and that of the proposed architecture in Section~\ref{s-experiment}.

The proposed scheme differs from the popular DDR DRAM interface in that the proposed architecture does not require an additional memory clock, since REB is replaced by the bidirectional DVS signal. Replacing REB by DVS, rather than adding an extra pin, is beneficial for maintaining backward compatibility with conventional components and boards.

Note that in the proposed architecture we rename WEB as RWEB, since it is used for both read and write modes.
%


\subsection{Proposed Interface Architecture}

Fig.~\ref{fig4} shows the block diagram of the proposed DDR synchronous NAND flash memory interface. As stated early, REB has been replaced by DVS for synchronous operations, and the FIFOs and latches have been duplicated for DDR operations.
The multiplexers are used inside the NAND flash chip in order to select WLAT for writes and RLAT for reads, depending on the edge type of RWEB. Now that RWEB is commonly used for both read and write modes, we do not need to distinguish $t_{WC}$ and $t_{RC}$ and thus use $t_{RWC}$ as the common timing parameter representing $t_{WC}$ and $t_{RC}$.
The D\_CON and Gen\_R blocks are not required in the proposed interface design but are included in the design shown in Fig.~\ref{fig4} for guaranteeing backward compatibility.

Note that the timing-critical path in the read mode is broken into two parts in the proposed design. One is the path for propagating RWEB, and the other is the data path from the NAND flash memory to the controller.
The delay of the first path determines $t_{RWC}$, since RWEB propagates through the same path in the write mode.
Thus, $t_{RWC}$ is identical to $t_{WC}$, rather than $t_{RC}$ of the conventional NAND flash memory.
The delay of the data path in the proposed architecture is shorter than $t_{RC}$ of the conventional architecture. This is because the propagation delay of RWEB does not need to be considered for calculating the data propagation delay. Consequently, the proposed interface can provide higher data throughput than the conventional one can.

To generate DVS at a stable data point, we use a delay-locked loop
(DLL) circuit. DLL is triggered by the data from RLAT and generates DVS by
delaying RWEB to satisfy the setup time ($t_{IOS}$) and the hold time
($t_{IOH}$) constraints at the input of the controller. We define
the time delay by the DLL as t$_{DLL}$, which is given by
\begin{equation} \label{eq1}
    t_{DLL} = t_{IOD,\mathrm{max}} - t_{RWEBD,\mathrm{min}} + t_{IOS}
\end{equation}
\noindent
where $t_{RWEBD}$ is the propagation delay of RWEB from the input
port of the NAND flash memory to the DLL, and $t_{IOD}$ is the data propagation delay from RLAT to the IO pads of the NAND flash memory. Note that the small variation in data availability can easily be adjusted by the DLL block.

\begin{figure*}
\centering
\subfigure[Write mode] {
    \includegraphics[width=0.67\linewidth]{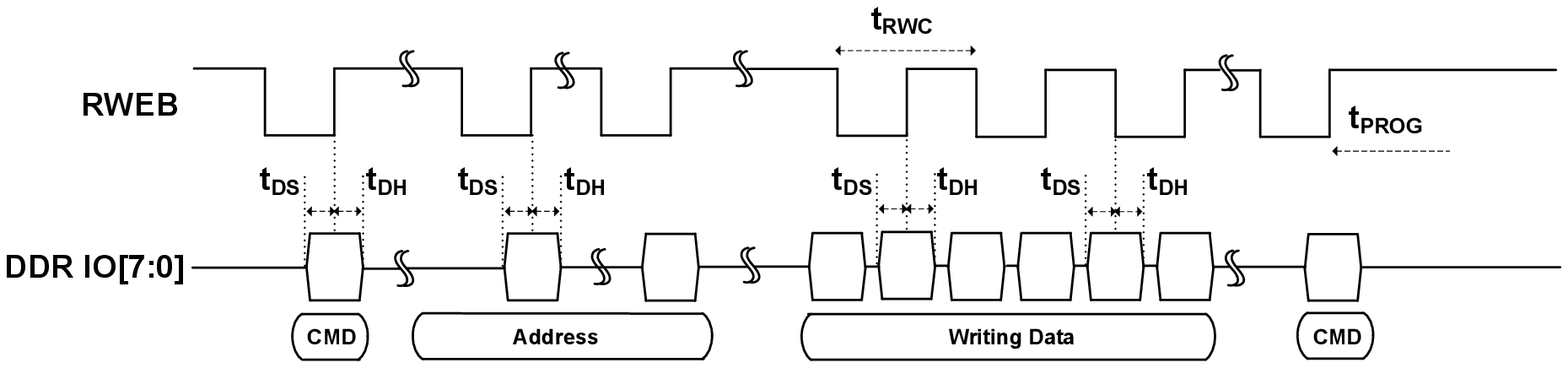}
}
\subfigure[Read mode] {
    \includegraphics[width=0.67\linewidth]{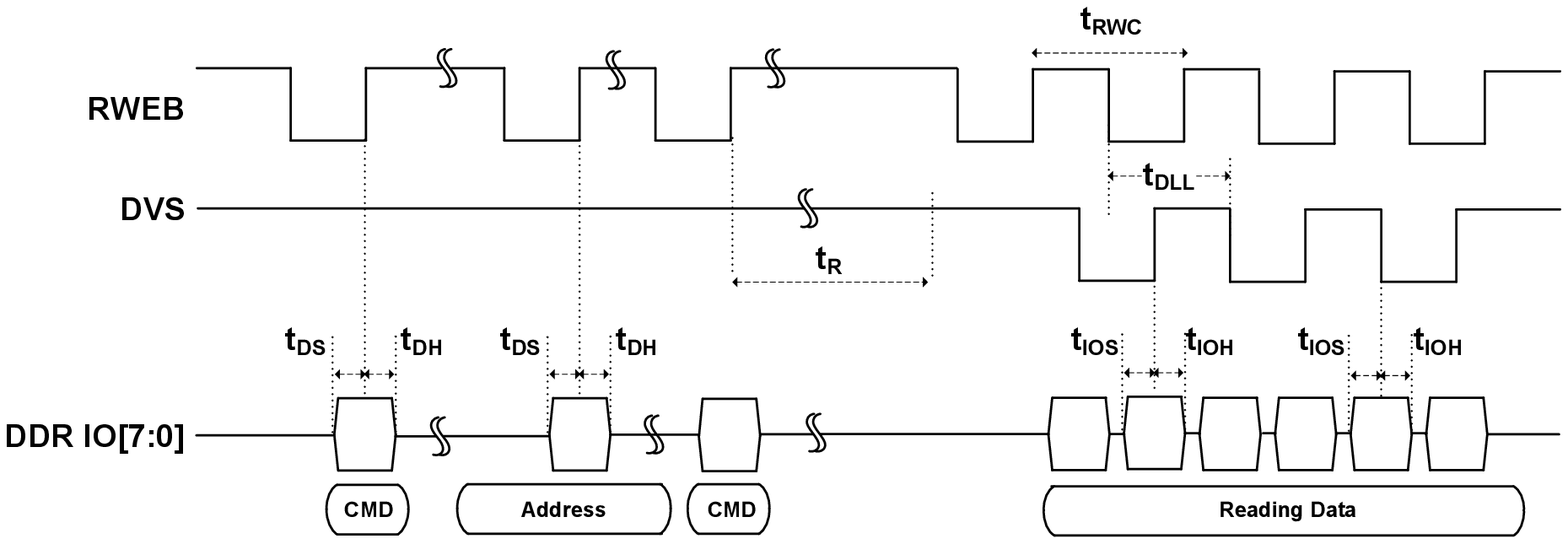}
}
\caption{Timing diagrams of the proposed DDR synchronous NAND flash interface.} \label{fig5}
\end{figure*}

\subsection{Write/Read Operation and Timing}

Fig.~\ref{fig5} shows the write and read timing diagrams of the
proposed DDR synchronous NAND flash interface. In the proposed interface, data is transferred at both rising and falling edges of the RWEB signal in the write mode, as represented in Fig.~\ref{fig5}(a). The data transfer rate can thus be improved by a factor of two compared with the conventional design.
In the read mode shown in Fig.~\ref{fig5} (b), the controller
asserts RWEB, instead of REB, to the NAND flash memory at $t_R$ after issuing the second CMD is completed. At the same time, the first data is pre-fetched to RLAT from the page register. The data are then moved from RLAT to the
IO ports and the DLL block that delays RWEB by $t_{DLL}$ for DVS
generation. Finally, the controller fetches the data at the
falling edge of DVS. For the next series of data, DVS is generated
in a similar manner, and the controller fetches at both edges of
DVS.

The major difference of the proposed design with respect to
the conventional one is the concurrent propagation of control
signals and data. Hence, it is possible for the proposed scheme to
have a shorter read cycle than the conventional design.

\subsection{Determining Operating Clock Period}
To compare the proposed and the conventional architectures in terms of their maximum operating frequency, we calculate the minimum period of the system clock (i.e. $t_{P,\rm{min}}$) for each architecture.

\subsubsection{Conventional Interface}

By design, $t_P$ should be at least the larger of $t_{RC}$ and $t_{WC}$, which are the periods of REB and WEB, respectively. From Section~\ref{ss-conv-reading} recall that $t_{RC} > t_{WC}$ since the propagation of REB and data should be serialized and happen within the same cycle in the read mode. Thus, we can ignore $t_{WC}$ for computing $t_{P,\rm{min}}$.

To determine $t_{P, \rm{min}}$, we also need to consider $t_{BYTE}$ since the data transfer between the page register and RLAT occurs in a distinct clock cycle that precedes the REB and data propagation. If this $t_{BYTE}$ parameter is greater than $t_{RC}$, $t_{P,\rm{min}}$ should be determined by $t_{BYTE}$. Consequently, $t_{P,\rm{min}}$ is given by
\begin{equation}\label{e-tp-min-conv}
t_{P,\rm{min}} = \max\{t_{RC}, t_{BYTE}\}.
\end{equation}

Since RFIFO is clocked by D\_CON, which delays CLK by $t_D$, the propagation of REB and data can take longer than $t_{RC}$, as already explained in Section~\ref{ss-conv-reading}. In other words, the following equality should hold:
\begin{equation}\label{e-trc-td}
t_{RC}+t_D = \underbrace{t_{OUT}}_{\textrm{For~REB}} + \underbrace{t_{REA} + t_{IN} + t_S}_{\textrm{For~data}}.
\end{equation}
Plugging Eq.~(\ref{e-trc-td}) into Eq.~(\ref{e-tp-min-conv}) gives
\begin{equation}\label{e-zzz}
t_{P,\rm{min}} = \max\{t_{OUT} + (t_{REA}+t_{IN}+t_{S})-t_{D}, t_{BYTE}\}
\end{equation}
which further develops to
\begin{equation}\label{e-conv-final}
t_{P,\rm{min}}=\max\left\{\frac{t_{OUT} + (t_{REA}+t_{IN}+t_{S})}{1+\alpha} ,t_{BYTE}\right\},
\end{equation}
by applying Eq.~(\ref{e-td}) to Eq.~(\ref{e-zzz}). The maximum clock frequency of the conventional design can then be determined by Eq.~(\ref{e-conv-final}).

\subsubsection{Proposed Interface}

For the proposed architecture, the value of $t_{P}$ should be at least the larger of $t_{RWC}$ and $t_{BYTE}$, namely
\begin{equation}
t_{P,\rm{min}} = \max\{t_{RWC}, t_{BYTE}\},
\end{equation}
since $t_{RWC}$ plays the role of $t_{RC}$.

\begin{figure}
\centering
\subfigure[] {
    \includegraphics[width=0.6\linewidth]{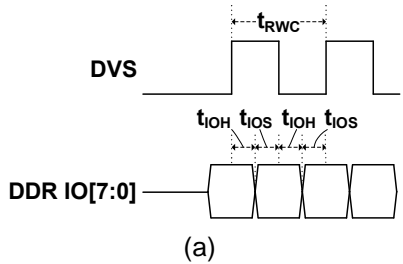}
}
\subfigure[] {
    \includegraphics[width=0.9\linewidth]{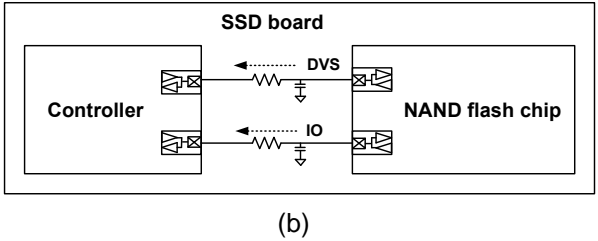}
}
\caption{Determining the minimum clock period of the proposed architecture: (a) $t_{RWC}$ should be at least $(t_{IOS}+t_{IOH})\times 2$. (b) The interconnect delays for DVS and IO are different.}
\label{fig-timing}
\end{figure}

Recall that the parameters $t_{IOS}$ and $t_{IOH}$ represent the setup and hold time constraints of data with respect to DVS at the IO pad of the controller, respectively. By design, $t_{RWC}$ is identical to the period of DVS, which should be at least twice the sum of $t_{IOS}$ and $t_{IOH}$, as shown in Fig.~\ref{fig-timing}(a). In other words. $t_{P,\rm{min}}$ of the proposed architecture is given by
\begin{equation}\label{eq3}
t_{P,\rm{min}} = \max\{(t_{IOS} + t_{IOH}) \times 2, t_{BYTE}\},
\end{equation}
where the term $(t_{IOS} + t_{IOH})$ is doubled since the proposed design supports DDR, and a single DVS cycle should thus be long enough to manage two transfers.
%

The architecture shown in Fig.~\ref{fig4} assumes that the controller and the NAND flash memory chips are integrated into a single board. Thus, $t_{IOS}$ and $t_{IOH}$ are affected by the geometric parameters of the board-level interconnects. When the board-level design parameters are available, we can derive an alternative representation of $t_{P,\rm{min}}$ given by
\begin{equation}\label{eq4}
t_{P,\rm{min}} = \max\left\{(t_{S} + t_{H} + t_{DIFF})\times 2, t_{BYTE}\right\},
\end{equation}
where $t_S$ and $t_H$ are the setup and hold times of RFIFO, respectively, and $t_{DIFF}$ is the difference between the arrival time of DVS to RFIFO and the arrival time of IO in the NAND flash memory to RFIFO. As informally shown in Fig.~\ref{fig-timing}(b), $t_{DIFF}$ is caused by the different interconnect delays of DVS and IO at the board level. In Eq.~(\ref{eq4}), note that $t_S$ and $t_H$ are independent of the geometric parameters of the board and that $t_{DIFF}$ also becomes a constant once the geometric parameters of the interconnects at the board have been decided.

The maximum clock frequency of the proposed design can be determined from either Eq.~(\ref{eq3}) or Eq.~(\ref{eq4}).

\section{Experimental Results} \label{s-experiment}
We present our results obtained from the experiments conducted to evaluate the performance of the new interface architecture proposed in Section~\ref{s-ddr}. In particular, we measured the write and read bandwidths of various SSD architectures that utilize the proposed interface design but are based upon different architectural and device-level choices such as the amount of channels, the degree of way interleaving and the type of flash cell (i.e. SLC/MLC). In addition, we measured the energy consumption of the proposed architecture. By comparison with the conventional flash interface, we show how much impact the proposed scheme has on the SSD-level performance in a variety of scenarios. 



After detailing the experimental setup in Section~\ref{ss-exp-setup}, we explain in Section~\ref{ss-exp-aux} how the operating frequencies of the tested architectures were determined. Section~\ref{ss-exp-ssd} presents the results of our experiments conducted to evaluate the read/write performance and energy consumption at SSD-level.



\subsection{Experimental Setting}\label{ss-exp-setup}


Based upon the basic architecture shown in Fig.~\ref{fig1}, two versions of SSD simulators were implemented: one for the conventional design and the other for the proposed design. The former employs the asynchronous interface shown in Fig.~\ref{fig2}, whereas the latter utilizes the DDR synchronous interface depicted in Fig.~\ref{fig4}. The controllers in both simulators were synthesized with the library built on a 130-nanometer process technology. The worst-case condition of this library consists of the IO voltage of 2.7 volts (V), the internal voltage of 1.35 V, and the temperature of 125 $^\circ$C. The timing parameters of the controllers shown in Fig.~\ref{fig2} and Fig.~\ref{fig4} were extracted using Synopsys PrimeTime$^\circledR$~\cite{primetime}.



The NAND flash memory simulated in the experiments was modeled at behavioral level with the timing parameters specified in~\cite{r18} and~\cite{r19} for SLC and MLC implementations, respectively, except for $t_{BYTE}$. Choosing a reasonable value of $t_{BYTE}$ is crucial for realistic simulation results since the maximum data transfer rate may be directly determined by $t_{BYTE}$ as shown in Eqs.~(\ref{e-conv-final}) and (\ref{eq4}). If the value of $t_{BYTE}$ is too high, then the first terms in these equations are eclipsed by $t_{BYTE}$ due to the $\max\{\cdot\}$ operator. For our experiments, the value of $t_{BYTE}$ was chosen from~\cite{r20}, which contains the specifications of OneNAND, one of the fastest (i.e. of the smallest $t_{BYTE}$) NAND flash memory commercially available. Note that the conventional NAND flash memory chips such as OneNAND are fabricated with only a single metal layer due to cost issues. If an additional metal layer is used, $t_{BYTE}$ would decrease further, and the performance gap between the proposed and the conventional architectures would become wider.

For the workload used in the experiments, we used widely used sequential traces that consist of 64-KB read/write data chunks~\cite{r21}. The sequential traces represent the typical access patterns happening when a large volume of data is written to or read from a storage based on NAND flash memory. As host interface, the SATA interface\footnote{We used SATA2 or ``SATA 3 Gbit/s,'' which supports the bandwidth of up to 300 MB/s.} was used. Finally, the overall SSD system was modeled at behavior level, and all the aforementioned models were integrated using MentorGraphics Seamless~\cite{seamless}.



\subsection{Operating Frequency Determination}\label{ss-exp-aux}
Using the simulators we developed, the major timing parameters of the proposed and the conventional interface architectures were measured, as listed in Table~\ref{t1}. The value of $t_{DIFF}$ was measured using CubicWare~\cite{r22,r23}; the difference of the loading capacitances of DVS and IO at the board set to 30 pF.
The values of $t_{S}$ and $t_{H}$ are identical for both architectures since they were synthesized with the same library. Note that only the first five parameters in the table were obtained from measurements; the rest are from the specification of NAND flash chips~\cite{r18,r19,r20}.


\begin{table}
\renewcommand{\arraystretch}{1.3}
\caption{NAND flash memory timing parameter values used in the experiments.}
\label{t1} \centering
\begin{tabular}{|c||c|c|}
\hline
 Parameters & Conventional (ns) & Proposed (ns)\\
\hline\hline
$t_{OUT}$ & 7.82 & N/A \\
\hline
 $t_{IN}$ & 1.65 & N/A \\
\hline
 $t_{S}$ & 0.25 & 0.25 \\
\hline
 $t_{H}$ & 0.02 & 0.02 \\
\hline
 $t_{DIFF}$ & N/A & 4.69 \\
\hline
 $t_{REA}$ & 20 & N/A \\
\hline
 t$_{BYTE}$ & 12 & 12 \\
\hline
\end{tabular}
\end{table}

For the conventional SSD, the minimum data access period $t_{P,\rm{min}}$ defined in Eq.~(\ref{e-conv-final}) can be evaluated as $t_{P,\rm{min}}=\max\left\{\frac{7.82+20+1.65+0.25}{1+0.5}, 12\right\}=19.81$ nanoseconds (ns) with the value of $\alpha=0.5$. Based on this, the maximum data access rate of the conventional design was set to 50 MHz. %
For the proposed design, Eq.~(\ref{eq4}) is evaluated as $t_{P,\rm{min}} = \max\{0.25+0.2+4.69,12\} = 12$ ns, and the maximum data access rate of the proposed design was set to 83 MHz.

%
%
%
%
%
%


\subsection{SSD-Level Performance Analysis}\label{ss-exp-ssd}
We compared and contrasted the performance of the SSDs designed with the proposed synchronous DDR interface with that of the SSDs using the conventional interface. The comparison criteria used were i) the write and read speeds, which have become one of the most important performance metrics for comparing different SSDs, and ii) energy consumption. 

%



Throughout the two sets of experiments detailed in Sections~\ref{ss-multi-way} and \ref{ss-multi-ch}, we wanted to see how the proposed architecture can guide the design decisions about the internal channel architecture; this is critical since it can trade-off between the area and performance of the SSD under design.

Three different interface designs were implemented and compared: the conventional asynchronous interface outlined in Section~\ref{s-conventional}, the synchronous (but not double-data-rate) interface proposed in~\cite{r17} and the proposed synchronous double-data-rate interface explained in Section~\ref{s-ddr}. In this section, these designs are referred to as CONV, SYNC\_ONLY and PROPOSED, respectively.

For convenience in implementation, the SYNC\_ONLY architecture was not developed from the scratch but was derived from PROPOSED by replacing DDR transfers with single-data-rate transfers. The operating frequency of SYNC\_ONLY was thus set to 83 MHz.


\subsubsection{Architectures with Different Way Interleaving}\label{ss-multi-way}

\begin{figure}
\centering
\subfigure[Single-Level Cell] {
    \includegraphics[width=\linewidth]{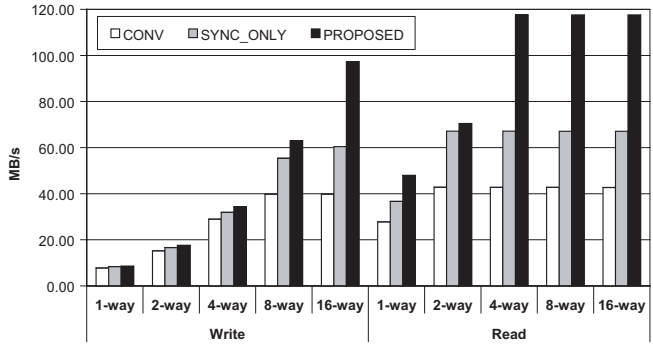}
}
\subfigure[Multi-Level Cell] {
    \includegraphics[width=\linewidth]{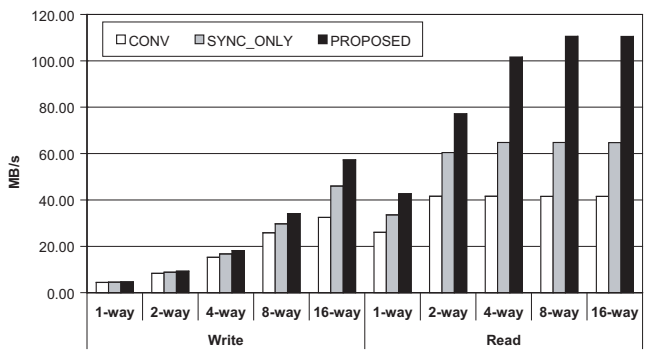}
}
\caption{Write/read speed of single-channel SSDs designed with different degrees of way interleaving (see Table~\ref{t-way-performance} for more details).} \label{fig7-8}
\end{figure}

We designed single-channel SSDs with five different degrees of way interleaving: 1-way, 2-way, 4-way, 8-way and 16-way. The write and read performance of each design was then measured for the three competing interfaces and the two flash cell types, as shown in Fig.~\ref{fig7-8} and Table~\ref{t-way-performance}. The experimental results we obtained clearly indicate that the proposed design greatly improves the system performance in corporation with the way-interleaving technique, as detailed below.

\begin{table}
\centering
\renewcommand{\arraystretch}{1.3}
\caption{Details of the values drawn in Fig.~\ref{fig7-8}.}
\label{t-way-performance}

\begin{minipage}{\linewidth}
\centering
\begin{tabular}{|c|c|c|c|c|c|c|c|}
\hline
\multirow{2}{*}{Cell} & \multirow{2}{*}{Mode} & \multirow{2}{*}{Way} & \multicolumn{3}{c|}{Performance (MB/s)} & \multicolumn{2}{c|}{Ratio}\\
\cline{4-8}
 &  &  & C$^{\dagger}$ & S & P & P/S & P/C\\
\hline\hline
\multirow{12}{*}{SLC} & \multirow{6}{*}{Write} & 1 & 7.77  & 8.38  & 8.50  & 1.01  & 1.09 \\
\cline{3-8}
 &  & 2 & 15.22  & 16.59  & 17.52  & 1.06  & 1.15 \\
\cline{3-8}
 &  & 4 & 28.94  & 31.90  & 34.30  & 1.08  & 1.19 \\
\cline{3-8}
 &  & 8 & 39.78  & 55.36  & 63.00  & 1.14  & 1.58 \\
\cline{3-8}
 &  & 16 & 39.76  & 60.44  & 97.35  & 1.61  & \textbf{2.45} \\
\cline{3-8}
 &  & \emph{Mean$^{\ddagger}$} & \emph{26.29}  & \emph{34.53}  & \emph{44.13}  & \emph{1.16}  & \emph{1.42} \\
\cline{2-8}
 & \multirow{6}{*}{Read} & 1 & 27.78  & 36.66  & 47.89  & 1.31  & 1.72 \\
\cline{3-8}
 &  & 2 & 42.78  & 67.16  & 70.47  & 1.05  & 1.65 \\
\cline{3-8}
 &  & 4 & 42.75  & 67.13  & 117.68  & 1.75  & 2.75 \\
\cline{3-8}
 &  & 8 & 42.72  & 67.11  & 117.64  & 1.75  & 2.75 \\
\cline{3-8}
 &  & 16 & 42.69  & 67.11  & 117.59  & 1.75  & \textbf{2.75} \\
\cline{3-8}
 &  & \emph{Mean} & \emph{39.74}  & \emph{61.03}  & \emph{94.25}  & \emph{1.49}  & \emph{2.26} \\
\hline\hline
\multirow{12}{*}{MLC} & \multirow{6}{*}{Write} & 1 & 4.43  & 4.55  & 4.65  & 1.02  & 1.05 \\
\cline{3-8}
 &  & 2 & 8.36  & 8.85  & 9.24  & 1.04  & 1.11 \\
\cline{3-8}
 &  & 4 & 15.24  & 16.75  & 18.13  & 1.08  & 1.19 \\
\cline{3-8}
 &  & 8 & 25.86  & 29.72  & 34.08  & 1.15  & 1.32 \\
\cline{3-8}
 &  & 16 & 32.45  & 45.99  & 57.23  & 1.24  & \textbf{1.76} \\
\cline{3-8}
 &  & \emph{Mean} & \emph{17.27}  & \emph{21.17}  & \emph{24.67}  & \emph{1.11}  & \emph{1.26} \\
\cline{2-8}
 & \multirow{6}{*}{Read} & 1 & 26.04  & 33.58  & 42.69  & 1.27  & 1.64 \\
\cline{3-8}
 &  & 2 & 41.59  & 60.41  & 77.19  & 1.28  & 1.86 \\
\cline{3-8}
 &  & 4 & 41.55  & 64.76  & 101.61  & 1.57  & 2.45 \\
\cline{3-8}
 &  & 8 & 41.52  & 64.75  & 110.56  & 1.71  & 2.66 \\
\cline{3-8}
 &  & 16 & 41.50  & 64.73  & 110.52  & 1.71  & \textbf{2.66} \\
\cline{3-8}
 &  &\emph{Mean} & \emph{38.44}  & \emph{57.65}  & \emph{88.51}  & \emph{1.49}  & \emph{2.21} \\
\hline
\end{tabular}
\begin{flushleft}
\footnotesize{$^\dagger$ C: CONV, S: SYNC\_ONLY, P: PROPOSED}\newline
\footnotesize{$^\ddagger$ The arithmetic mean for columns 4--6; the geometric mean for columns 7--8.}
\end{flushleft}

\end{minipage}

\end{table}

\noindent \textbf{$\bullet$ Case I (write, SLC): }
We first consider the SLC cases shown in Fig.~\ref{fig7-8}(a). For the 1-way design, the write performance of CONV and PROPOSED is similar, the latter being better only by 9\%. This marginal improvement originates from the fact that the data transfer time from the SSD controller to the NAND flash memory is much smaller than the cell program time $t_{PROG}$ of the NAND flash memory. What PROPOSED reduces is the data transfer time, rather than $t_{PROG}$. By Amdahls' law, the impact of reducing the data transfer time on the overall performance is therefore diminished by the dominant size of $t_{PROG}$.



However, as the degree of way interleaving is increased, the advantage of using PROPOSED becomes more evident. For CONV, the performance gain by way interleaving decreases as the number of ways increases, eventually being saturated at the 8-way design. In contrast, for PROPOSED, the interleaving effect was maintained throughout all the degrees of way interleaving. Note that CONV achieved only about 5x performance gain as the number of ways changed from 1 to 16, whereas the performance gain by PROPOSED was more than 11x under the same condition. For the 16-way design, PROPOSED outperformed CONV by 2.45 times. This difference is caused by the fact that PROPOSED enables the controller to put more data in a fixed amount of time (i.e. $t_{PROG}$) than CONV.

The performance of SYNC\_ONLY lied between those of CONV and PROPOSED, as expected from the fact that SYNC\_ONLY does not support double-data-rate data transfers.

\noindent \textbf{$\bullet$ Case II (read, SLC): }
This case is shown in the right-hand side of Fig.~\ref{fig7-8}(a). The overall performance of reading was higher than that of writing for all the three interfaces tested. By design, the way-interleaving technique can fully be effective during $t_R$ in the read mode, while it does not fully utilize $t_{PROG}$ in the write mode.
Even in this case, the way-interleaving technique is more effective to PROPOSED, since the performance of PROPOSED is saturated at the larger degree of way-interleaving compared to CONV.
Namely,  PROPOSED and CONV are saturated when the degrees of way interleaving are 4-way and 2-way, respectively.
The relative performance of PROPOSED over CONV in the read mode was also higher than that in the write mode for all degrees of way interleaving. For instance, PROPOSED outperformed CONV by a factor of 2.75 for the 16-way design.




\noindent \textbf{$\bullet$ Case III (write/read, MLC): }
Fig.~\ref{fig7-8}(b) shows the results for the MLC NAND flash memory design. The read time ($t_R$) and the program time ($t_{PROG}$) parameters of MLC devices are much larger than those of SLC devices. Thus, the effect of way interleaving on the overall performance decreases in MLC devices for the same degree of way interleaving. This reduction in the effectiveness of way interleaving is larger in the write mode than in the read mode, since $t_{PROG}$ is much larger than $t_R$. This result indicates that the proposed interface combined with the interleaving technique can be more effective for high-capacity storage devices that are composed of many MLC chips than for low-capacity storages. We can also deduce that the proposed design is more advantageous for storage devices with many low-density MLC chips than for storages with a small number of high-density MLC chips.


\begin{figure}[!t]
\centering
\subfigure[Single-Level Cell] {
    \includegraphics[width=\linewidth]{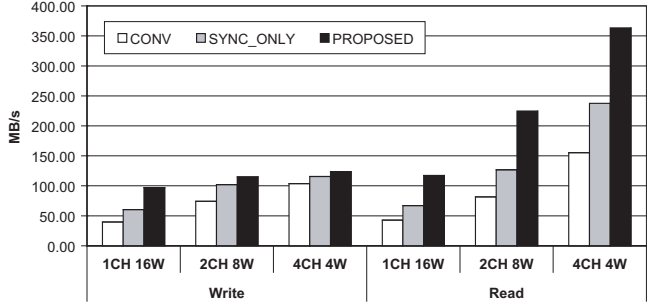}
}
\subfigure[Multi-Level Cell] {
    \includegraphics[width=\linewidth]{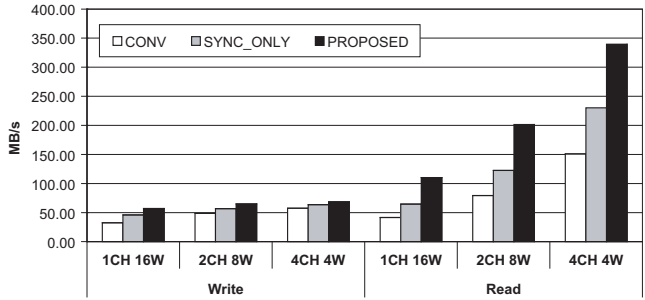}
}
\caption{Write/read speed of SSDs designed with different numbers of channels and degrees of way interleaving (see Table~\ref{t-vary-performance} for more details).} \label{fig10-11}
\end{figure}

\subsubsection{Architectures with Various Channel Configurations}\label{ss-multi-ch}
In practice, the capacity of a storage system is typically determined earlier than micro-architectural design parameters such as the number of ways and channels. Given a capacity value, we can explore the various combinations of ways and channels to search for optimal design. In this regard, we tested three different SSD architectures of varying channel/way configurations (1-channel/16-way, 2-channel/8-way and 4-channel/4-way), while keeping the product of channels and ways constant. In other words, the number of NAND flash chips (i.e. total capacity) used in each architecture was kept identical. Throughout this experiment, we wanted to determine the optimal number of channels and the degree of way interleaving, considering the trade-off between performance and area. For each design, the write and read speeds were measured both for SLC- and MLC-based implementations. The results are shown in Fig.~\ref{fig10-11} and Table~\ref{t-vary-performance}.




\begin{table}
\centering
\renewcommand{\arraystretch}{1.3}
\caption{Details of the values drawn in Fig.~\ref{fig10-11}.}
\label{t-vary-performance}

\begin{minipage}{\linewidth}
\centering
\begin{tabular}{|c|c|c|c|c|c|c|c|}
\hline
\multirow{2}{*}{Cell} & \multirow{2}{*}{Mode} & Ch- & \multicolumn{3}{c|}{Performance (MB/s)} & \multicolumn{2}{c|}{Ratio}\\
\cline{4-8}
 &  & Way & C$^{\dagger}$ & S & P & P/S & P/C\\
\hline\hline
\multirow{8}{*}{SLC} & \multirow{4}{*}{Write} & 1-16 & 39.76  & 60.44  & 97.35  & 1.61  & \textbf{2.45} \\
\cline{3-8}
 &  & 2-8 & 74.07  & 101.99  & 114.83  & 1.13  & 1.55 \\
\cline{3-8}
 &  & 4-4 & 103.76  & 115.68  & 123.52  & 1.07  & 1.19 \\
\cline{3-8}
 &  & \emph{Mean$^{\ddagger}$} & \emph{72.53}  & \emph{92.70}  & \emph{111.90}  & \emph{1.25}  & \emph{1.65} \\
\cline{2-8}
 & \multirow{4}{*}{Read} & 1-16 & 42.69  & 67.11  & 117.59  & 1.75  & 2.75 \\
\cline{3-8}
 &  & 2-8 & 81.44  & 126.70  & 224.82  & 1.77  & \textbf{2.76} \\
\cline{3-8}
 &  & 4-4 & 155.35  & 237.61  & max$^\S$  & --  & -- \\
\cline{3-8}
 &  & \emph{Mean} & \emph{93.16}  & \emph{143.81}  & \emph{235.25}  & \emph{1.76}  & \emph{2.76} \\
\hline\hline
\multirow{8}{*}{MLC} & \multirow{4}{*}{Write} & 1-16 & 32.45  & 45.99  & 57.23  & 1.24  & \textbf{1.76} \\
\cline{3-8}
 &  & 2-8 & 48.72  & 56.83  & 64.75  & 1.14  & 1.33 \\
\cline{3-8}
 &  & 4-4 & 57.46  & 63.55  & 68.49  & 1.08  & 1.19 \\
\cline{3-8}
 &  & \emph{Mean} & \emph{46.21}  & \emph{55.46}  & \emph{63.49}  & \emph{1.15}  & \emph{1.41} \\
\cline{2-8}
 & \multirow{4}{*}{Read} & 1-16 & 41.50  & 64.73  & 110.52  & 1.71  & \textbf{2.66} \\
\cline{3-8}
 &  & 2-8 & 79.32  & 122.48  & 201.42  & 1.64  & 2.54 \\
\cline{3-8}
 &  & 4-4 & 150.94  & 230.17  & max  & --  & -- \\
\cline{3-8}
 &  & \emph{Mean} & \emph{90.59}  & \emph{139.13}  & \emph{217.18}  & \emph{1.68}  & \emph{2.60} \\
\hline
\end{tabular}
\begin{flushleft}
\footnotesize{$^\dagger$ C: CONV, S: SYNC\_ONLY, P: PROPOSED}\newline
\footnotesize{$^\ddagger$ The arithmetic mean for columns 4--6; the geometric mean for columns 7--8.}\newline
\footnotesize{$^\S$ Reached the maximum bandwidth of the SATA interface.}
\end{flushleft}

\end{minipage}
\end{table}

\noindent \textbf{$\bullet$ Case I (write, SLC): }
In the write mode shown in Fig.~\ref{fig10-11}(a), the performance of PROPOSED increased more slowly than that of CONV as the area (i.e. the number of channels) increased. In our experiment, the architectures designed with more channels have fewer degrees of way interleaving, and thus the benefits of using PROPOSED decreases as more channels are used. In the write mode, it would therefore be better to increase the degree of way interleaving than to increase the number of channels if a tight area budget is given.

%
\noindent \textbf{$\bullet$ Case II (read, SLC): }
Unlike the write mode, the performance of the three interfaces increases in an almost identical fashion as more channels and fewer degrees of way interleaving are used. This is because the interval to which the way-interleaving technique is applied is much shorter in the read mode (i.e. $t_{R}$ in the read mode versus $t_{PROG}$ in the write mode). Note that the read bandwidth is much higher than the write bandwidth. Thus, the read bandwidth of the (4-channel, 4-way) configuration in Fig.~\ref{fig10-11}(a) actually reached the bandwidth of the SATA host interface we used.



\noindent \textbf{$\bullet$ Case III (write/read, MLC)}
Fig.~\ref{fig10-11}(b) shows the result from simulating MLC-based SSD designs. The overall performance pattern is similar to that appearing in Fig.~\ref{fig10-11}(a). However, the degree of performance improvements is smaller than that in the SLC case. For instance, in the SLC-based design, the read bandwidth of PROPOSED was improved by 1.91 times as the configuration changes from (1-channel, 16-way) to (2-channel, 8-way). In contrast, in the MLC-based scheme, the read performance of PROPOSED increases only by 1.81 times for the same change in channel and way configuration.

This phenomenon becomes more evident in the write mode. This is again related to the length of the period to which the way-interleaving technique can be applied. This period in the write mode is $t_{PROG}$, which is much larger than the counterpart $t_{R}$ in the read mode. In the write mode, a larger degree of way interleaving is required in order to  saturate the channel bandwidth. Thus, increasing channels is in effect only when the degree of way interleaving is sufficiently large. Typically, the difference in $t_{PROG}$ between MLC and SLC is much larger than the difference in $t_R$ between MLC and SLC. Therefore, the performance degradation of MLC-based SSDs is more clearly seen in the write mode.

\subsubsection{Energy Consumption Comparison}\label{ss-exp-power}

To see the impact of the proposed architecture on energy consumption, we first measured the average power consumption of the SSD controllers that adopt different interfaces, when these controllers read or write the same amount of data. Note that the operating frequencies of CONV, SYNC\_ONLY and PROPOSED are different. Thus, for fair comparison, we further divided the power consumption of an interface by the bandwidth (measured in megabytes per second) this interface operates at. In other words, we compared the energy consumed by the SSD controllers to transfer a single byte of data.

\begin{figure}
\centering
\includegraphics[width=\linewidth]{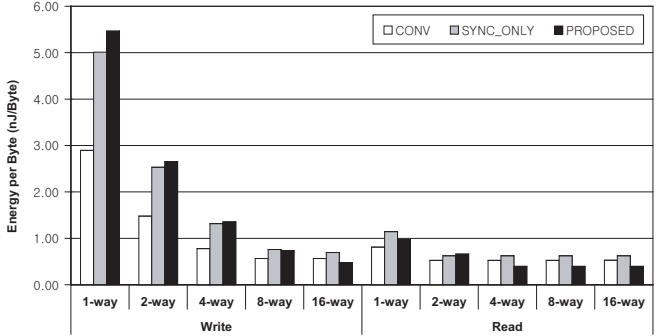}
\caption{Energy consumed by different SSD controllers to transfer a single byte (see Table~\ref{t-exp-power} for more details). Unit: nano-Joule per byte.} \label{fig12}
\end{figure}

Fig~\ref{fig12} and Table~\ref{t-exp-power} show the result we obtained from simulating the SLC-based designs for write and read operations for various degrees of way interleaving. For low degrees of way interleaving, PROPOSED consumed more energy than CONV to read or write the same amount of data. However, as the degree of way interleaving increases, the energy consumed by PROPOSED gradually became the smallest among the alternatives. Due to the performance issues, as discussed in Section~\ref{ss-multi-way}, it is likely that most SSDs will continue to be designed with a reasonably high degree of way interleaving. For such design, adopting the proposed interface would be highly beneficial, since it outperforms the alternatives not only in terms of the read/write bandwidth but also with respect to the energy efficiency.

\begin{table}
\centering
\renewcommand{\arraystretch}{1.3}
\caption{Details of the values drawn in Fig.~\ref{fig12}.}
\label{t-exp-power}

\begin{minipage}{\linewidth}
\centering
\begin{tabular}{|c|c|c|c|c|c|c|c|}
\hline
\multirow{2}{*}{Cell} & \multirow{2}{*}{Mode} & \multirow{2}{*}{Way} & \multicolumn{3}{c|}{Energy (nJ/B) } & \multicolumn{2}{c|}{Ratio}\\
\cline{4-8}
 &  &  & C$^{\dagger}$ & S & P & P/S & P/C\\
\hline\hline
\multirow{12}{*}{SLC} & \multirow{6}{*}{Write} & 1 & 2.90  & 5.01  & 5.47  & 1.09  & 1.89 \\
\cline{3-8}
 &  & 2 & 1.48  & 2.53  & 2.65  & 1.05  & 1.80 \\
\cline{3-8}
 &  & 4 & 0.78  & 1.32  & 1.36  & 1.03  & 1.74 \\
\cline{3-8}
 &  & 8 & 0.57  & 0.76  & 0.74  & 0.97  & 1.30 \\
\cline{3-8}
 &  & 16 & 0.57  & 0.69  & 0.48  & 0.69  & 0.84 \\
\cline{3-8}
 &  & \emph{Mean$^{\ddagger}$} & \emph{1.26}  & \emph{2.06}  & \emph{2.14}  & \emph{0.95}  & \emph{1.45} \\
\cline{2-8}
 & \multirow{6}{*}{Read} & 1 & 0.81  & 1.15  & 0.97  & 0.85  & 1.20 \\
\cline{3-8}
 &  & 2 & 0.53  & 0.63  & 0.66  & 1.06  & 1.25 \\
\cline{3-8}
 &  & 4 & 0.53  & 0.63  & 0.40  & 0.63  & 0.75 \\
\cline{3-8}
 &  & 8 & 0.53  & 0.63  & 0.40  & 0.63  & 0.75 \\
\cline{3-8}
 &  & 16 & 0.53  & 0.63  & 0.40  & 0.63  & 0.75 \\
\cline{3-8}
 &  & \emph{Mean} & \emph{0.58}  & \emph{0.73}  & \emph{0.56}  & \emph{0.74}  & \emph{0.91} \\
\hline
\end{tabular}
\begin{flushleft}
\footnotesize{$^\dagger$ C: CONV, S: SYNC\_ONLY, P: PROPOSED}\newline
\footnotesize{$^\ddagger$ The arithmetic mean for columns 4--6; the geometric mean for columns 7--8.}
\end{flushleft}

\end{minipage}
\end{table}




\section{Conclusion} \label{s-conclusion}
We have proposed a novel SSD architecture that exploits double-data-rate synchronous NAND flash interface. This new design not only enhances the write and read performance but also retains the backward compatibility with existing single-data-rate asynchronous NAND flash memory. The performance of the SSDs that exploit the way-interleaving technique can be greatly improved by adopting the proposed approach. Our experimental results show that the proposed architecture outperforms the conventional one by 1.65--2.76 times in the read mode and 1.09--2.45 times in the write mode for the SLC-architectures we considered. For the MLC-based architectures tested, the new design we propose improves the performance by 1.64--2.66 times in the read mode and 1.05--1.76 times in the write mode over the conventional design. The proposed scheme can dramatically increase the operating frequency of the interface, only limited by $t_{BYTE}$, which is the device-level parameter characterizes the read time of a flash cell. As process technology advances, $t_{BYTE}$ will keep decreasing, and the impact of our scheme will become more prominent.

\section*{Acknowledgment}
This work was supported by Samsung Electronics Co., Ltd., by IC Design Education Center (IDEC), by a KOSEF grant funded by the Korean Government (MEST)(No. 2009-0079888) and by a KRF grant funded by the Korean Government (MOEHRD) (No. KRF-2007-313-D00578).

\balance
\bibliographystyle{IEEEtran}
\bibliography{ddr}

\begin{thebibliography}{10}
\providecommand{\url}[1]{#1}
\csname url@samestyle\endcsname
\providecommand{\newblock}{\relax}
\providecommand{\bibinfo}[2]{#2}
\providecommand{\BIBentrySTDinterwordspacing}{\spaceskip=0pt\relax}
\providecommand{\BIBentryALTinterwordstretchfactor}{4}
\providecommand{\BIBentryALTinterwordspacing}{\spaceskip=\fontdimen2\font plus
\BIBentryALTinterwordstretchfactor\fontdimen3\font minus
  \fontdimen4\font\relax}
\providecommand{\BIBforeignlanguage}[2]{{%
\expandafter\ifx\csname l@#1\endcsname\relax
\typeout{** WARNING: IEEEtran.bst: No hyphenation pattern has been}%
\typeout{** loaded for the language `#1'. Using the pattern for}%
\typeout{** the default language instead.}%
\else
\language=\csname l@#1\endcsname
\fi
#2}}
\providecommand{\BIBdecl}{\relax}
\BIBdecl

\bibitem{katz1989dsa}
R.~Katz, G.~Gibson, and D.~Patterson, ``{Disk system architectures for high
  performance computing},'' \emph{Proceedings of the IEEE}, vol.~77, no.~12,
  pp. 1842--1858, 1989.

\bibitem{r1}
Y.~Choi, K.~Suh, Y.~Koh, J.~Park, K.~Lee, Y.~Cho, and B.~Suh, ``{A high speed
  programming scheme for multi-level NAND flash memory},'' in \emph{Proceedings
  of Symposium on VLSI Circuits}, 1996, pp. 170--171.

\bibitem{r2}
J.~Kim, K.~Sakui, S.~Lee, Y.~Itoh, S.~Kwon, K.~Kanazawa, K.~Lee, H.~Nakamura,
  K.~Kim, T.~Himeno \emph{et~al.}, ``{A 120-mm 2 64-Mb NAND flash memory
  achieving 180 ns/Byteeffective program speed},'' \emph{IEEE Journal of
  Solid-State Circuits}, vol.~32, no.~5, pp. 670--680, 1997.

\bibitem{r3}
K.~Takeuchi and T.~Tanaka, ``{A dual-page programming scheme for high-speed
  multigigabit-scale NAND flash memories},'' \emph{IEEE Journal of Solid-State
  Circuits}, vol.~36, no.~5, pp. 744--751, 2001.

\bibitem{r4}
J.~Lee, H.~Im, D.~Byeon, K.~Lee, D.~Chae, K.~Lee, S.~Hwang, S.~Lee, Y.~Lim,
  J.~Lee \emph{et~al.}, ``{High-performance 1-Gb-NAND flash memory with 0.
  12-mum technology},'' \emph{IEEE Journal of Solid-State Circuits}, vol.~37,
  no.~11, pp. 1502--1509, 2002.

\bibitem{r5}
K.~Imamiya, H.~Nakamura, T.~Himeno, T.~Yarnamura, T.~Ikehashi, K.~Takeuchi,
  K.~Kanda, K.~Hosono, T.~Futatsuyama, K.~Kawai \emph{et~al.}, ``{A 125-mm/sup
  2/1-Gb NAND flash memory with 10-MByte/s program speed},'' \emph{IEEE Journal
  of Solid-State Circuits}, vol.~37, no.~11, pp. 1493--1501, 2002.

\bibitem{r6}
T.~Hara, K.~Fukuda, K.~Kanazawa, N.~Shibata, K.~Hosono, H.~Maejima,
  M.~Nakagawa, T.~Abe, M.~Kojima, M.~Fujiu \emph{et~al.}, ``{A 146-mm 2 8-Gb
  multi-level NAND flash memory with 70-nm CMOS technology},'' \emph{IEEE
  Journal of Solid-State Circuits}, vol.~41, no.~1, pp. 161--169, 2006.

\bibitem{r7}
K.~Takeuchi, Y.~Kameda, S.~Fujimura, H.~Otake, K.~Hosono, H.~Shiga,
  Y.~Watanabe, and T.~Futatsuyama, ``{A 56-nm CMOS 99-Formula Not Shown 8-Gb
  Multi-Level NAND Flash Memory With 10-MB/s Program Throughput},'' \emph{IEEE
  Journal of Solid-State Circuits}, vol.~42, no.~1, p. 219, 2007.

\bibitem{r11}
J.~Kim, J.~Kim, S.~Noh, S.~Min, and Y.~Cho, ``{A space-efficient flash
  translation layer for compactflash systems},'' \emph{IEEE Transactions on
  Consumer Electronics}, vol.~48, no.~2, pp. 366--375, 2002.

\bibitem{r12}
S.~Kim and S.~Jung, ``{A log-based flash translation layer for large NAND flash
  memory},'' in \emph{Proceedings of the 8th International Conference (ICACT
  2006)}, 2006, pp. 1641--1644.

\bibitem{r13}
C.~Wu and T.~Kuo, ``{An adaptive two-level management for the flash translation
  layer in embedded systems},'' in \emph{Proceedings of the 2006 IEEE/ACM
  international conference on Computer-aided design}.\hskip 1em plus 0.5em
  minus 0.4em\relax ACM New York, NY, USA, 2006, pp. 601--606.

\bibitem{bragdon1995fmm}
M.~Assar, S.~Nemazie, P.~Estakhri \emph{et~al.}, ``{Flash memory mass storage
  architecture incorporation wear leveling technique},'' Dec.~26 1995, united
  States Patent 5,479,638.

\bibitem{sata}
\url{http://www.serialata.org}.

\bibitem{kim2001ssd}
D.~Kim, K.~Bang, S.~Ha, C.~Park, S.~Chung, and E.~Chung, ``{Solid-State Disk
  with Double Data Rate DRAM Interface for High-Performance PCs},'' \emph{IEICE
  Trans. on Information and Systems}, vol. E92-D, no.~4, pp. 727--731, 2009.

\bibitem{r16}
C.~Park, P.~Talawar, D.~Won, M.~Jung, J.~Im, S.~Kim, and Y.~Choi, ``{A high
  performance controller for NAND Flash-based solid state disk (NSSD)},'' in
  \emph{Proceedings of the 21st Non-Volatile Semiconductor Memory Workshop
  (IEEE NVSMW)}, 2006, pp. 17--20.

\bibitem{cache1}
D.~Ryu, ``{Solid state disk controller apparatus},'' Dec.~19 2005, united
  States Patent App. 11/311,990.

\bibitem{cache2}
J.~Lee and D.~Ryu, ``{Semiconductor solid state disk controller},'' Nov.~9
  2006, united States Patent App. 11/594,893.

\bibitem{r8}
K.~Yim, H.~Bahn, and K.~Koh, ``{A flash compression layer for SmartMedia card
  systems},'' \emph{IEEE Transactions on consumer Electronics}, vol.~50, no.~1,
  pp. 192--197, 2004.

\bibitem{r9}
W.~Huang, C.~Chen, Y.~Chen, and C.~Chen, ``{A compression layer for NAND type
  flash memory systems},'' in \emph{Proceedings of the Third International
  Conference on Information Technology and Applications (ICITA 2005)}, vol.~1,
  2005.

\bibitem{r10}
W.~Huang, C.~Chen, and C.~Chen, ``{The Real-Time Compression Layer for Flash
  Memory in Mobile Multimedia Devices},'' in \emph{Proceedings of International
  Conference on Multimedia and Ubiquitous Engineering (MUE'07)}, 2007, pp.
  171--176.

\bibitem{r14}
L.~Chang and T.~Kuo, ``{An adaptive striping architecture for flash memory
  storage systems of embedded systems},'' in \emph{Proceedings of the Eighth
  IEEE Real-Time and Embedded Technology and Applications Symposium}, 2002, pp.
  187--196.

\bibitem{r15}
S.~Lim and K.~Park, ``{An efficient NAND flash file system for flash memory
  storage},'' \emph{IEEE Transactions on Computers}, pp. 906--912, 2006.

\bibitem{r17}
C.~Son, S.~Yoon, S.~Chung, C.~Park, and E.~Chung, ``{Variability-insensitive
  scheme for NAND flash memory interfaces},'' \emph{Electronics Letters},
  vol.~42, no.~23, pp. 1335--1336, 2006.

\bibitem{onfi}
\url{http://www.onfi.org}.

\bibitem{hland}
R.~Schuetz, H.~Oh, J.~Kim, H.~Pyeon, S.~Przybylski, and P.~Gillingham,
  ``{Hyperlink nand flash architecture for mass storage applications},'' in
  \emph{Proceedings of IEEE Non-Volatile Semiconductor Memory Workshop}, 2007,
  pp. 3--4.

\bibitem{r18}
\emph{K9F1G08U0B 128M x 8-bit NAND Flash Memory Data Sheet V1.0}, Samsung
  Electronics Company, 2006.

\bibitem{r19}
\emph{K9GAG08U0M 2G x 8-bit NAND Flash Memory Data Sheet V1.0}, Samsung
  Electronics Company, 2006.

\bibitem{r20}
\emph{FK8G16Q2M 2G 2Gb MuxOneNAND M-die Data Sheet V1.1}, Samsung Electronics
  Company, 2007.

\bibitem{primetime}
\url{http://www.synopsys.com/Tools/Implementation/SignOff/Pages/PrimeTime.aspx}.

\bibitem{r21}
\emph{MultiMediaCard System Specification Version 4.2}, MMCA MultiMediaCard
  Association, 2006.

\bibitem{seamless}
\url{http://http://www.mentor.com/products/fv/seamless/}.

\bibitem{r22}
M.~Jang, H.~Jin, B.~Lee, J.~Lee, S.~Song, T.~Kim, and J.~Kong, ``{CubicWare: a
  hierarchical design system for deep submicron ASIC},'' in \emph{In
  Proceedings of the Twelfth Annual IEEE International ASIC/SOC Conference},
  1999, pp. 168--172.

\bibitem{r23}
\url{http://www.samsung.com/global/business/semiconductor/products/asic/Products_EDASupport.html}.

\end{thebibliography}
\end{document}